\documentclass[prd,aps,floatfix,nofootinbib,preprint,tightenlines]{revtex4}
\usepackage{dcolumn}
\usepackage{amsmath}
\usepackage{latexsym}
\usepackage{graphicx}
\usepackage{bm}

\def\svev#1{\left\langle #1\right\rangle}       

\long \def \blockcomment #1\endcomment{}

\newcommand{\bee}{\begin{equation}}
\newcommand{\ee}{\end{equation}}
\newcommand{\beea}{\begin{eqnarray}}
\newcommand{\eea}{\end{eqnarray}}

\begin{document}
\title{
Log-normal distribution for correlators in lattice QCD?}
\author{Thomas DeGrand}%
 \affiliation{Department of Physics,
University of Colorado, Boulder, CO 80309, USA}

\begin{abstract}
Many hadronic correlators  used in spectroscopy calculations in lattice QCD
simulations appear to show a log-normal distribution at intermediate time separations. 
\end{abstract}

\maketitle

\begin{figure}
\begin{center}
\includegraphics[width=0.8\textwidth,clip]{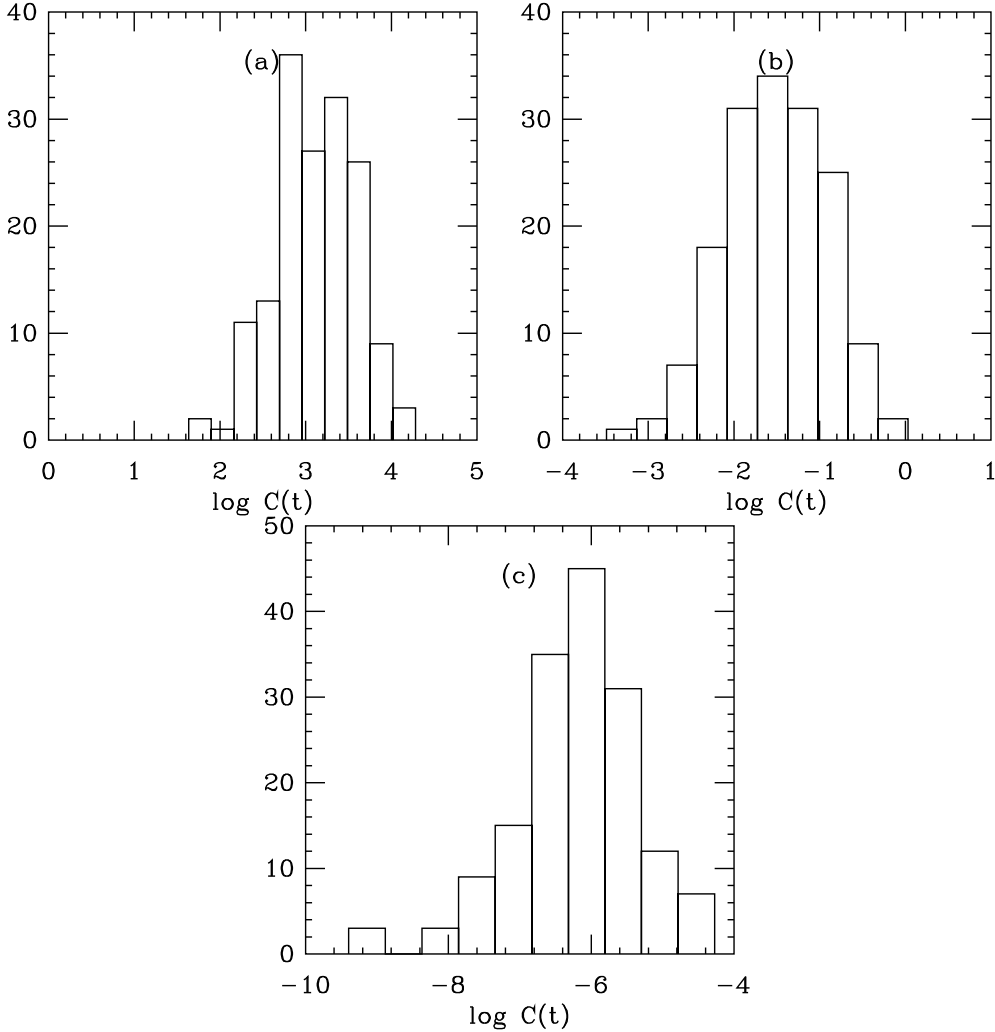}
\end{center}
\caption{Histogram of values of $\log C(t)$ for the propagator of a $J=7/2$ baryon in
$SU(7)$. Panels (a), (b), and (c) show results for $t=4$, 6, and 8 respectively.
\label{fig:SU7delt2}}
\end{figure}

Recently, while performing numerical simulations of unitary fermion gases, the authors of 
Refs.~\cite{Endres:2011mm,Lee:2011sm,Endres:2011er,Endres:2011jm,Endres:2010sq,Nicholson:2010ms,Lee:2010qp}
 discovered that  
 spectroscopic correlation functions   of operators separated by 
a Euclidean time $t$, call them generically $C(t)$,
show a log-normal distribution.
In a review article \cite{Endres:2011mm}, they 
presented high statistics plots of the distribution
of propagator values 
of a Lambda-Lambda dibaryon state\cite{Detmold}, which also show a beautiful Gaussian structure for $\log C(t)$.
The width of the Gaussian increases roughly linearly with $t$. I have been doing
simulations of quenched baryon spectroscopy in larger-$N$ $SU(N)$ gauge field backgrounds,
and I see the same thing, although with
 lower statistics: compare Fig.~\ref{fig:SU7delt2}.

Unitary Fermi gases, Lambda-Lambda dibaryons, and large-N baryons are rather exotic objects
 for lattice study,
and the question naturally arises, how common are log-normal distributions
 in lattice spectroscopy? I believe
that they are ubiquitous. I observe them in the following data sets:
\begin{itemize}
\item Meson and baryon spectroscopy, and string tension data from Wilson loops,
in quenched $SU(3)$, $SU(5)$ and $SU(7)$ simulations at a lattice spacing of about 0.1 fm
\item $N_f=2$ flavor dynamical simulations at a similar lattice spacing
\item Quenched $SU(3)$ simulations with the Wilson action at $\beta=5.9$ and 6.1 and overlap valence fermions
\item Simulations in the weak coupling phase of $SU(3)$ gauge theory with two flavors of
sextet-representation fermions
\end{itemize}
This is a qualitative observation.
 I do not know why it occurs, how general it might be, nor what it is good for.
In order not to make the paper too long, and to avoid being too
redundant, I will only show pictures from quenched QCD. 

Let's set some definitions. With the $n$th moment of a set of
random variables $x_i$ ($i=1$ to $N$)
as
\bee
\mu_n=\svev{x^n},
\ee
the $n$th cumulant $\kappa_n$ is defined recursively as
\bee
\kappa_n =  \mu_n - \sum_{m=1}^{n-1} \left(
   \begin{array}{c} n-1 \\ m-1 \end{array} \right) \kappa_m \mu_{n-m}.
\ee
The objects of our attention are some  set of expectation values of correlation functions of pairs of operators $O$
\bee
C(t) = \sum_x O_l(x,t) O_m(0,0) 
\ee
generated in a Monte Carlo simulation, a set of random variables,
 $C_i(t)$ for the $i$th measurement. Their falloff with $t$ gives mass values.

If the operators are built of fermion propagators (such as for a meson or baryon propagator),
lattice symmetries (charge conjugation plus $\gamma_5$ Hermiticity) tell us that the real part
of $C(t)$ carries the signal. In an infinite ensemble the imaginary part of $C(t)$ would average to zero.
So I will only consider sets of real variables $C_i(t)$.
The $x$'s will  be the logarithms of $C(t)$. At small and intermediate $t$
all the $C(t)$'s in any data set have the same sign (positive, by definition).
 At the largest times, some of
the $C(t)$'s in some correlation functions can fluctuate negative.
 When I calculate the cumulants of $\log C(t)$, I will simply discard these
wrong-sign entries from my analysis, and when I show a result I will report the number of
discarded configurations.

I consider mesonic and baryonic correlation functions, and Wilson loops.
The correlators I show which involve quark propagators use clover fermions with 
links  smeared
using normalized hypercubic (nHYP) smearing \cite{Hasenfratz:2001hp,Hasenfratz:2007rf}.
The clover
coefficient  is set to unity.
Most of my spectroscopic sets use an extended source (typically, a Gaussian product state
is used as the source of the fermion propagator) and point sinks,
 projected onto zero three momentum.
Some data sets use zero-momentum Gaussian sinks as well. The first class of correlators
is not variational. The size of the source has typically been tuned to produce flat plateaus 
in effective mass plots.
 These are completely standard data
 sets for lattice simulations,
although the largest lattice volumes are small by today's standards, $16^3\times 48$ sites.

The other set of correlators I consider are Wilson loops, used to compute the heavy quark potential.
These are real quantities and should all have the same sign. Fluctuations can drive them 
negative and I will treat this situation as I do for mesonic or baryonic correlators.
 The loops come from lattice configurations which were
nHYP smeared and gauge fixed to axial gauge.

I observe that over a wide range in $t$ the second cumulant $\kappa_2$ is much greater than
the higher ($n>2$) cumulants. If a distribution
is Gaussian, its  first and second cumulants (the mean and standard deviation) are the only
 non-vanishing ones, so this ordering of moments means that the distributions of $C(t)$
are approximately log-normal.
I also observe the same ordering of size of  the $m$th moment of the correlator $M_n$, defined as
a power of the original correlation function
\bee
M_n(t)= (C(t))^n .
\ee
Moments of a log-normally distributed variable are also log-normal.
Finally, I observe, like
Refs.~\cite{Endres:2011mm,Lee:2011sm,Endres:2011er,Endres:2011jm,Endres:2010sq,Nicholson:2010ms,Lee:2010qp},
that  the second cumulant of $\log C(t)$, $\kappa_2$, increases roughly linearly with
 time $t$.
 Log-normal behavior is
most prominent at short and intermediate distances,
 but these are distances where effective mass plots are roughly
constant, where one would take masses to publish as results.

\begin{figure}
\begin{center}
\includegraphics[width=0.9\textwidth,clip]{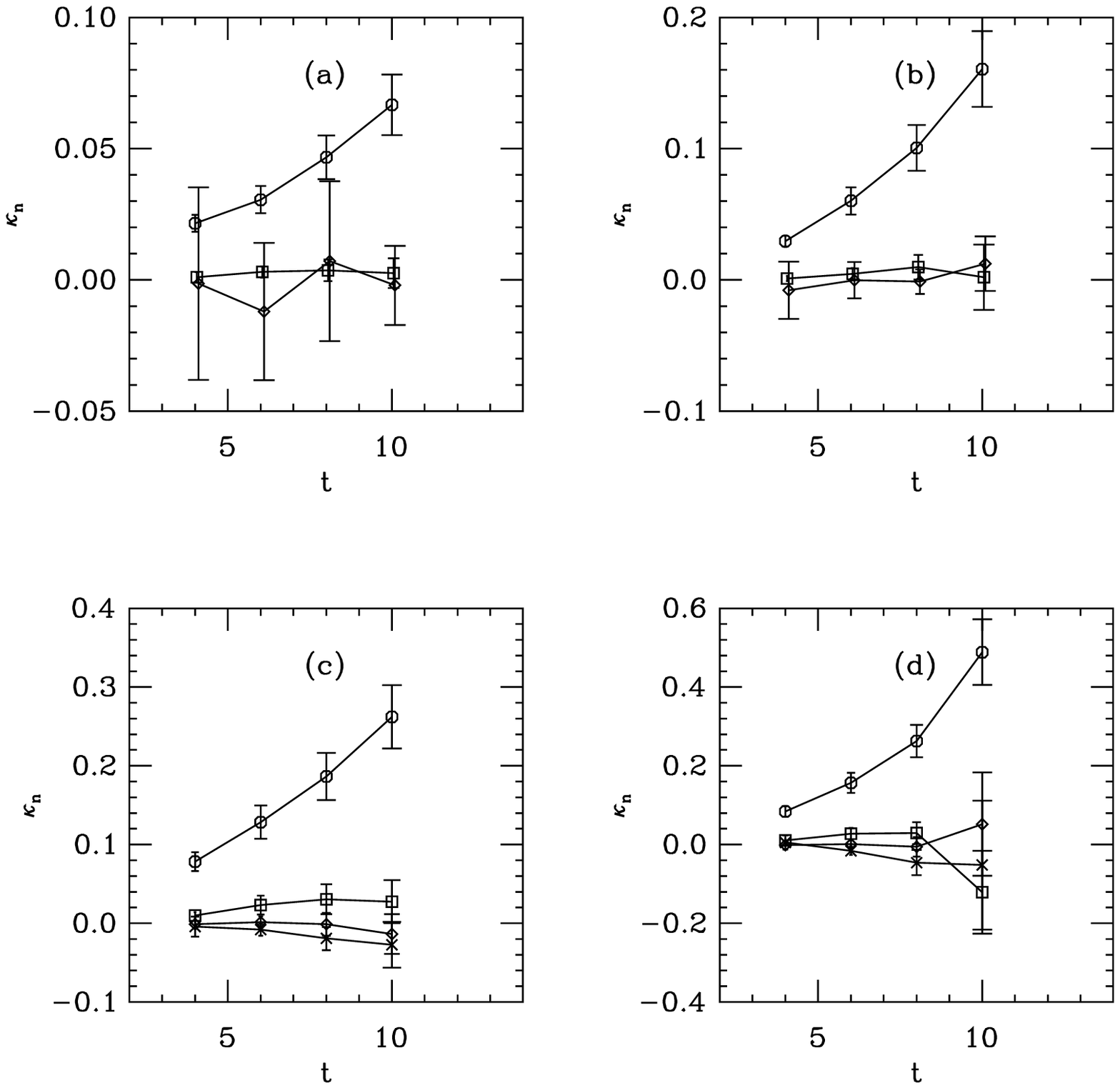}
\end{center}
\caption{Cumulants of $\log C(t)$ for various smeared-to-point hadronic correlators of temporal extent $t$,
from quenched $SU(3)$ simulations at $\beta=6.0175$, $\kappa=0.125$.
(a) pseudoscalar, (b) vector, (c) proton (d) Delta.
 Labels are octagons for $\kappa_2$, squares for $\kappa_3$, diamonds for $\kappa_4$,
crosses for $\kappa_5$. All correlators are positive at all $t$ apart from one Delta correlator at $t=10$.
\label{fig:had6.0175}}
\end{figure}

Let us look at some examples. I begin with a data
 set of 80 $16^3\times 32$ quenched $SU(3)$ lattices
at $\beta=6.0175$.
Hadronic correlators from this data set, at one $\kappa$ corresponding to an Axial Ward Identity (AWI) quark mass
in lattice units of
 about $am_q=0.055$,
are shown in Fig.~\ref{fig:had6.0175}.
 The errors on the $\kappa_n$'s come from a jackknife.
Observe that $\kappa_2$ is much larger than the other $\kappa_n$'s and increases linearly with $t$.

\begin{figure}
\begin{center}
\includegraphics[width=0.9\textwidth,clip]{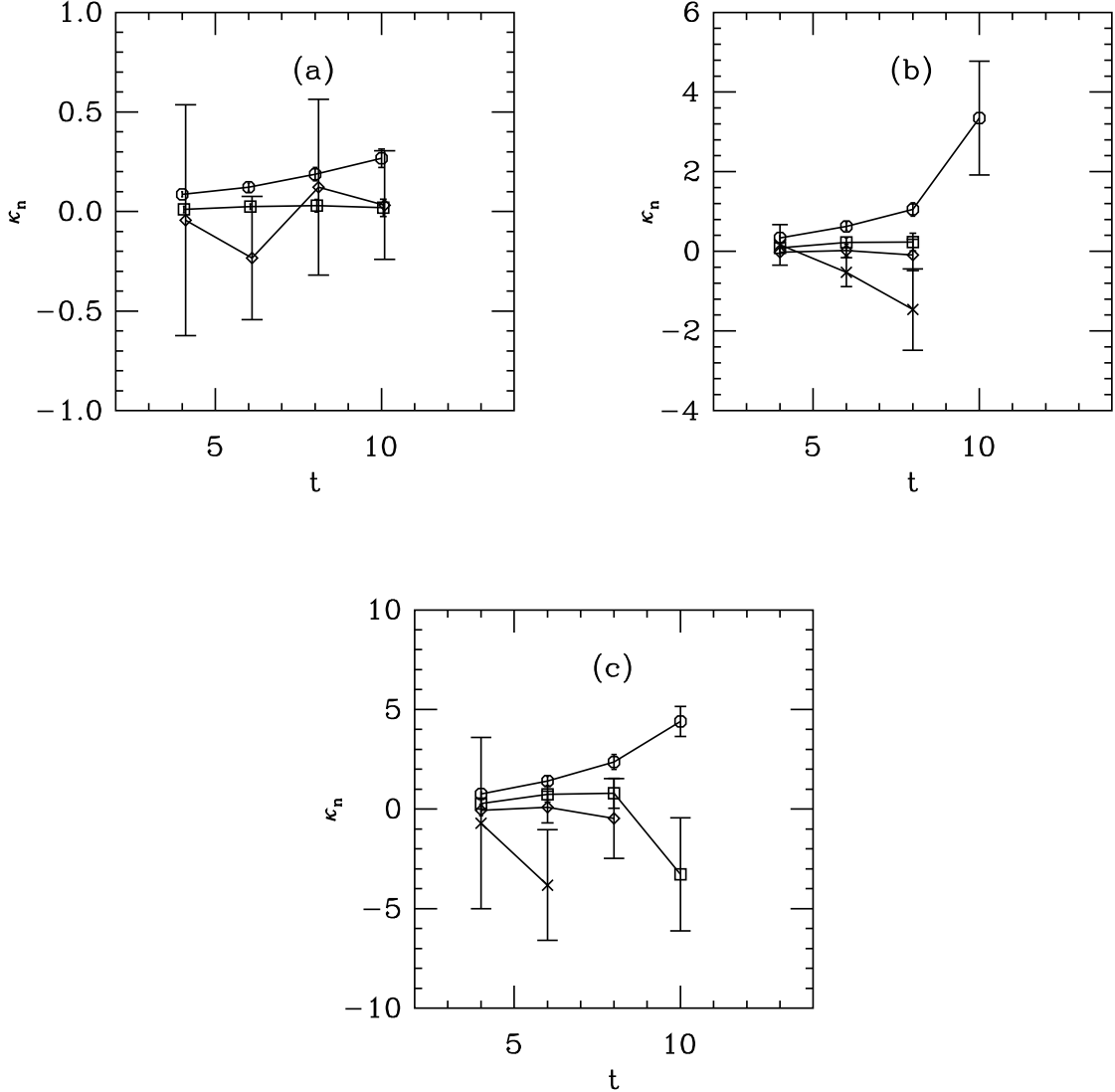}
\end{center}
\caption{Cumulants of $\log M_n(t)$ for moments of various smeared-to-point hadronic correlators of temporal extent $t$,
from quenched $SU(3)$ simulations at $\beta=6.0175$, $\kappa=0.125$.
(a) square of the pseudoscalar correlator, (b) square of the Delta (c)  cube of the Delta.
 Labels are octagons for $\kappa_2$, squares for $\kappa_3$, diamonds for $\kappa_4$,
crosses for $\kappa_5$ (baryons only; these are very noisy for the mesons). 
\label{fig:Nhad6.0175}}
\end{figure}

Fig.~\ref{fig:Nhad6.0175} shows a set of cumulants from moments. The
 higher cumulants become quite noisy.
The second moment of the square of the pseudoscalar
propagator, the square of the delta propagator, and the cube of the delta propagator all
 increase with $t$ and they
either dominate the other moments or remain the only cumulant with statistically
 significant signal.

\begin{figure}
\begin{center}
\includegraphics[width=0.9\textwidth,clip]{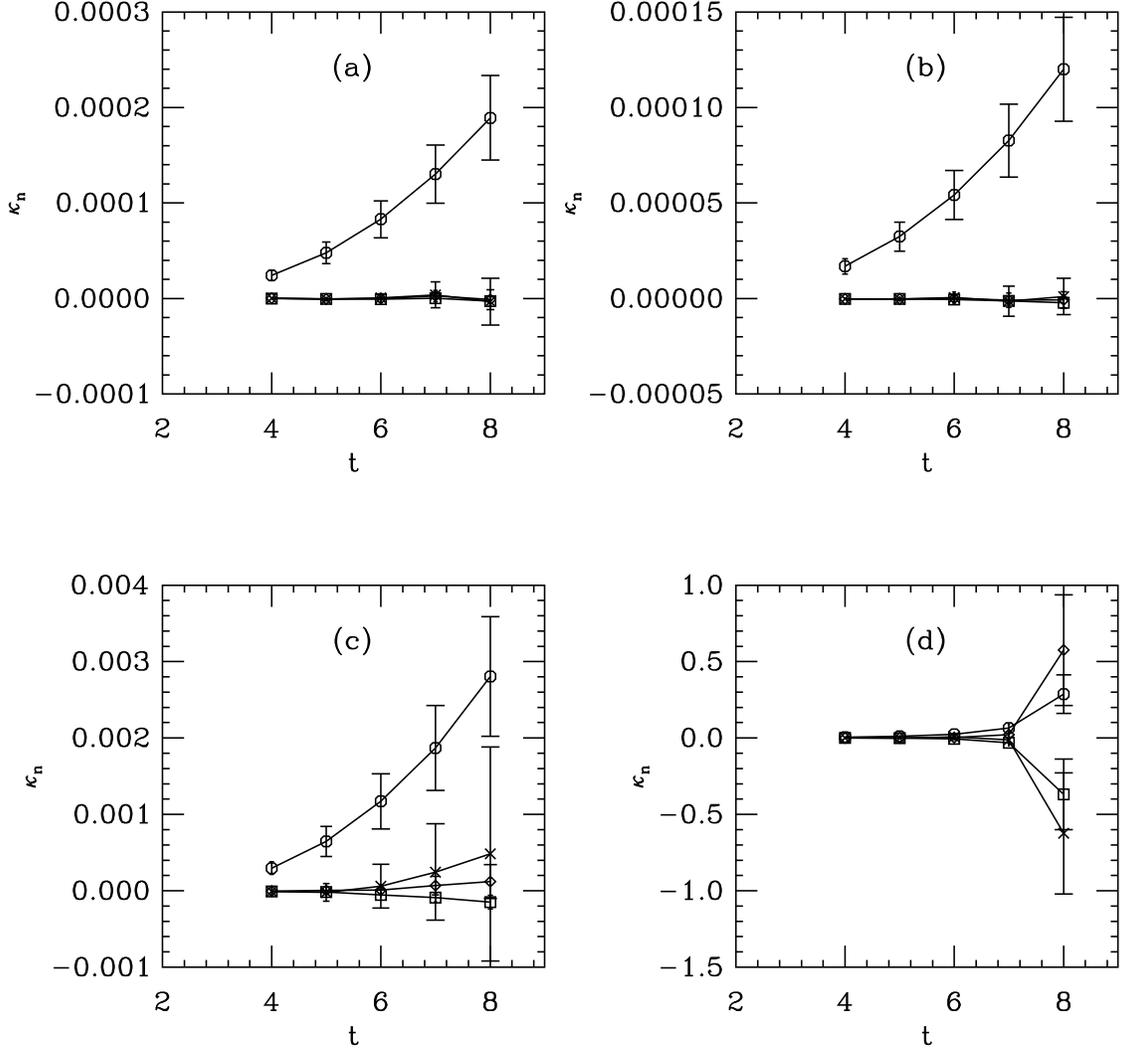}
\end{center}
\caption{Cumulants of $\log C(t)$ for various Wilson loops of temporal extent $t$,
from quenched $SU(3)$ simulations at $\beta=6.0175$.
(a) $r=2$ planar loop;
(b) $\vec r= (1,1,1)$ loop;
(c) $r=3\sqrt{2}$ loop;
(d) $r=6\sqrt{2}$ loop.
 Labels are octagons for $\kappa_2$, squares for $\kappa_3$, diamonds for $\kappa_4$,
crosses for $\kappa_5$. All correlators are positive at all $t$.
\label{fig:pot6.0175}}
\end{figure}

\begin{figure}
\begin{center}
\includegraphics[width=0.9\textwidth,clip]{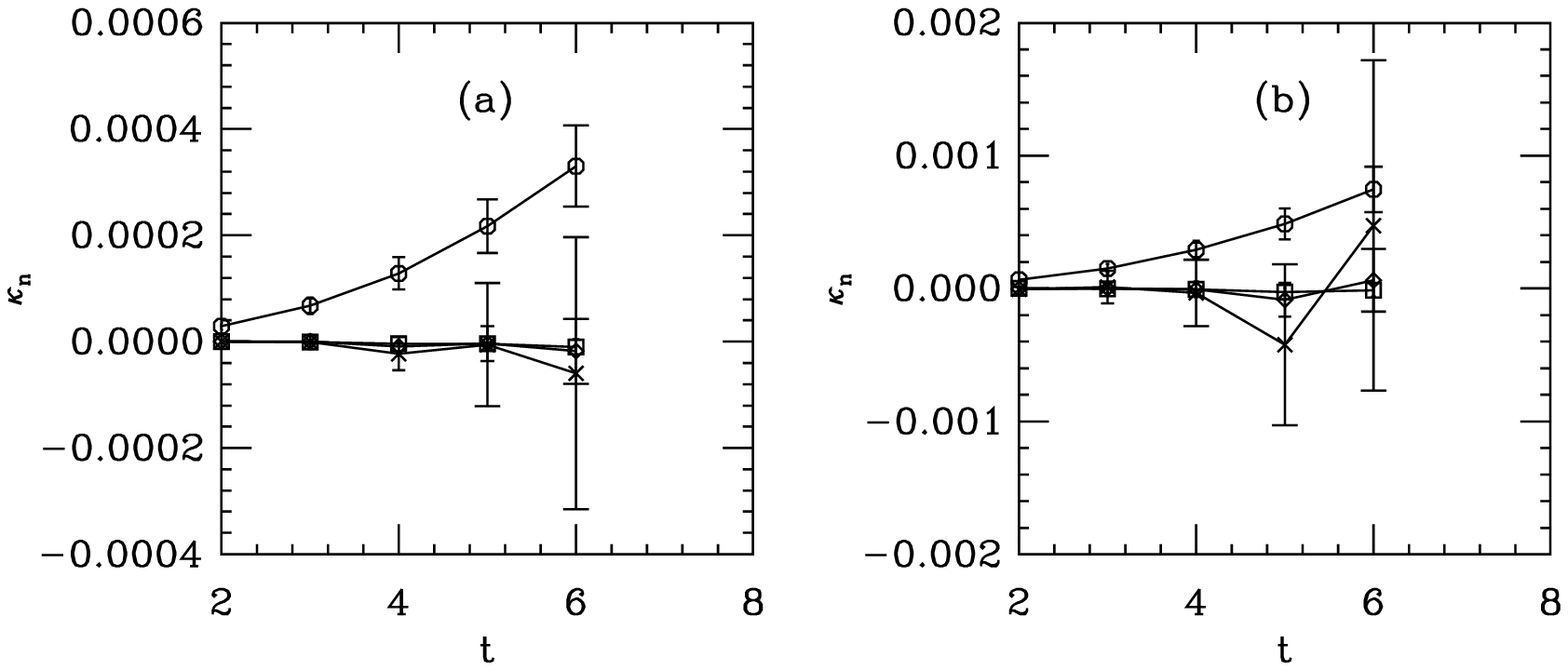}
\end{center}
\caption{Cumulants of $\log M_n(t)$ from quenched $SU(3)$ simulations at $\beta=6.0175$.
for the (a) second and (b) third moments of
the $r=(1,1,1)$ Wilson loop.
 Labels are octagons for $\kappa_2$, squares for $\kappa_3$, diamonds for $\kappa_4$,
crosses for $\kappa_5$. 
\label{fig:Nwl6.0175}}
\end{figure}

 Fig.~\ref{fig:pot6.0175} shows plots of $\kappa_n$ vs $t$ for
 several Wilson loops
of temporal extent $t$ from this data set.
Distances of $t=4-6$ are the range from which potentials may begin to be be reliably extracted.
Cumulants for the second and third moments of the $(1,1,1)$ loop (panel (b) of Fig.~\ref{fig:pot6.0175})
are shown in Fig.~\ref{fig:Nwl6.0175}. Evidently, these
Wilson loop expectation values are also log-normal distributed.

Log normal distributions present a contradiction with a well-known expectation of the noise
in correlators, which goes back to
Lepage\cite{Lepage:1989hd}. The (over) simplified version of the explanation is that while
the signal $C_H(t)$ decays as $\sim \exp(-m_H t)$,
the noise in the channel involves the exponential decay of the absolute square
of the correlator
\bee
\sigma^2(t) \sim |C(t)|^2 = \langle 0| |O(t)|^2 |O(0)|^2 | 0 \rangle \sim \exp(-m_2t)
\label{eq:lepage}
\ee
where $m_2$ is the lightest state which can be created by the squared operator.
For the pseudoscalar or the rho, the lightest state is the two-pseudoscalar state and
$\sigma(t)/C(t)$ should be roughly
a constant for the pseudoscalar, roughly increasing exponentially as $\exp((m_\rho-m_\pi))t$ for the rho.
(The energy of two particle states in a 
box includes an interaction term\cite{Luscher:1986pf}, which will reappear below.)
 For the ($N$ color) baryon correlator,
two different classes of behavior are expected for the moments\cite{savagefootnote}:
when the moment number
$n$ is even, the correlator should couple to $nN/2$ pseudoscalars and when $n$ is odd, the lightest
state will be a single baryon plus $(n-1)N/2$ pseudoscalars.
Sometimes the squared correlator  can couple to the vacuum,
in which case $\sigma^2(t)$ would be a constant. This is the situation for the scalar glueball mass
or any Wilson loop.

Now consider the situation for a log-normal correlator. The
average value of the $n$th moment is
\bee
\langle x^n\rangle = \exp(n \kappa_1 + \frac{n^2}{2}\kappa_2).
\label{eq:power}
\ee
 The
correlators $C(t)$  decrease with $t$ proportional to $\exp(-M t)$. This says that
both $\kappa_1$ and $\kappa_2$ should be linear functions of $t$, which is what we have seen.
Call
\bee
\kappa_2(t)= tS + S_0.
\label{eq:S}
\ee
We can define an effective mass from the logarithm of the ratio, 
\bee M=-\log \svev{C(t+1)}/\svev{C(t)}.
\label{eq:effmass}
\ee
Eq.~\ref{eq:power} tells us that the mass associated
with the $n$th moment is
\bee
M_n = nM_1 - \frac{n(n-1)}{2}S
\label{eq:twobody}
\ee
that is, the log-normal distribution implies a pairwise interaction of the constituents of the
$n$th moment. This is clearly inconsistent with Lepage-like behavior.

We can compare the two expectations for correlators.
 With the correlators in hand, just construct the correlation functions by averaging powers of the $C(t)$'s,
the $n$th moments, $M_n(t)$ and directly measure the effective mass of $M_n(t)$.

\begin{figure}
\begin{center}
\includegraphics[width=0.9\textwidth,clip]{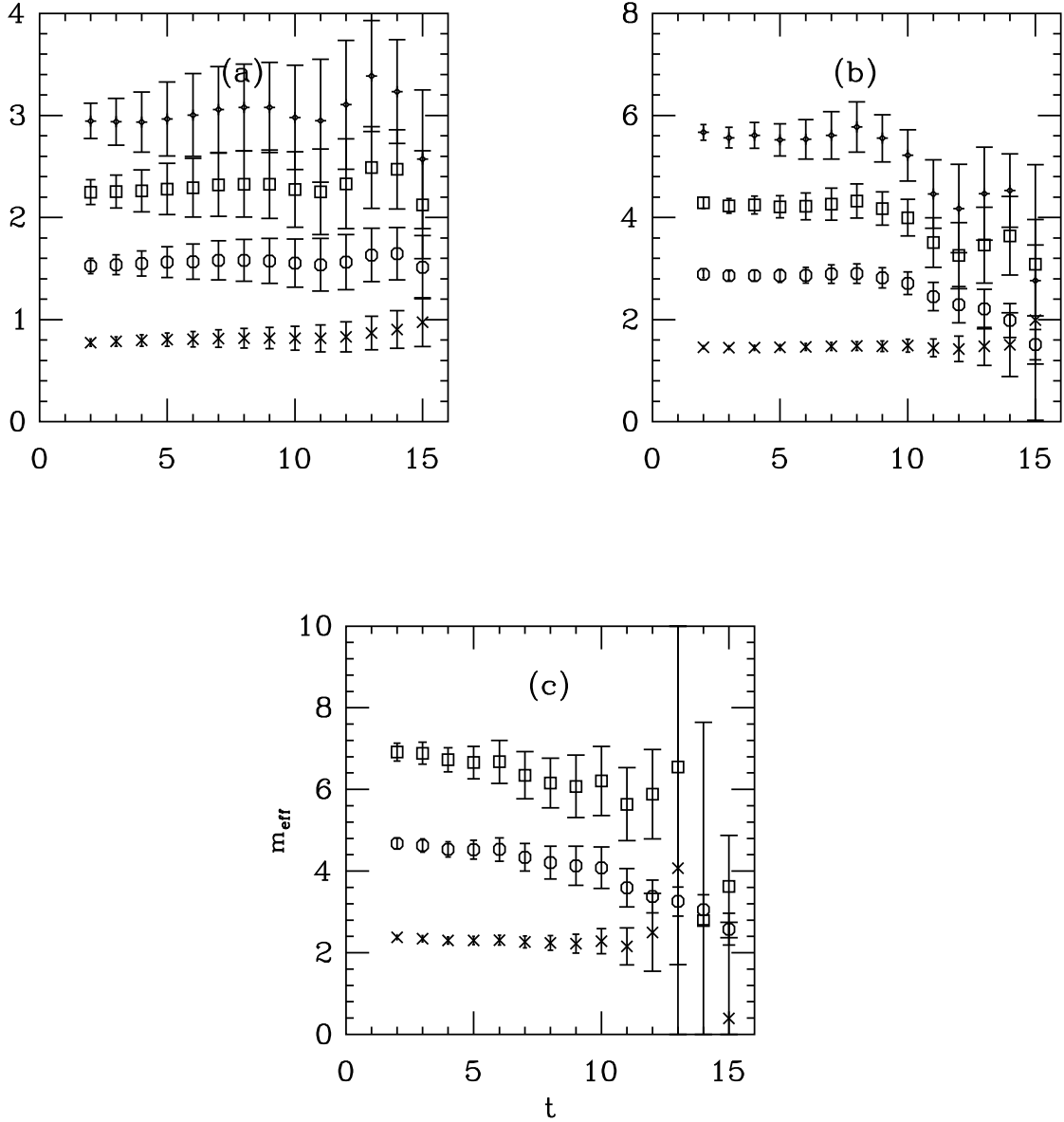}
\end{center}
\caption{Effective mass for moments of the highest-spin baryon (higher moments lie higher, so the
baryon effective mass is given by crosses, the effective mass of the squared correlator is given by octagons, 
for the cubed correlator, by squares, and the fancy diamonds are for $M_4$):
(a) $SU(3)$,  $\kappa=0.125$;
(b) $SU(5)$,  $\kappa=0.1265$;
(c) $SU(7)$,  $\kappa=0.128$.
\label{fig:moment}}
\end{figure}

In Fig.~\ref{fig:moment},  I compare the highest-spin baryon in $SU(N)$, $N=3$,
5, 7. The bare couplings have been tuned to match the lattice spacings.
  Panel (a) shows moments of the SU(3) delta.
The lightest state is just the delta itself: its mass (in lattice units) is about 0.8.
At its $\kappa$ value the lattice pseudoscalar mass is 0.35, so the second moment (the octagons)
should asymptote to a mass of $3\times 0.35=1.05$. Instead, it sits at roughly twice the delta's mass.
The second moment does not show a mass which is the sum of the delta mass plus 
 three times the pseudoscalar, 1.85; it sits
at roughly three times the delta mass.

For $SU(5)$ (panel (b)), the situation is similar. Again the lattice pseudoscalar mass is 0.35.
The baryon mass is about 1.5 in lattice units. The $n$th moment's effective mass is roughly just
$n$ times the baryon mass over a wide $t$ range. At large $t$ the masses tail over toward the Lepage formula.
This is a soft statement, because the quality of the fit has deteriorated and it may be that
the signal is just overwhelmed by noise, but it is certainly plausible.
Note that this behavior occurs at much larger $t$ than where 
the baryon's effective mass has gone to a plateau.

The situation for $SU(7)$ (panel (c)) is again similar. (Here, the pseudoscalar mass is 0.4 in lattice
units). Apparently
Eq.~\ref{eq:lepage} is only an asymptotic result. This is no surprise: the simple story was too simple.
 The correlator couples to
everything with its quantum numbers, not just the lightest state:
\bee
M_n(t) = \sum_j Z_j \exp(-m_j t)
\ee
where $m_j$ can include the $n-$ baryon state. Presumably this is a dominant state, since
some attempt was made to optimize the operators
 to produce a single baryon state in $C(t)$. So the asymptotic form 
may appear only at very late time.

\begin{figure}
\begin{center}
\includegraphics[width=0.9\textwidth,clip]{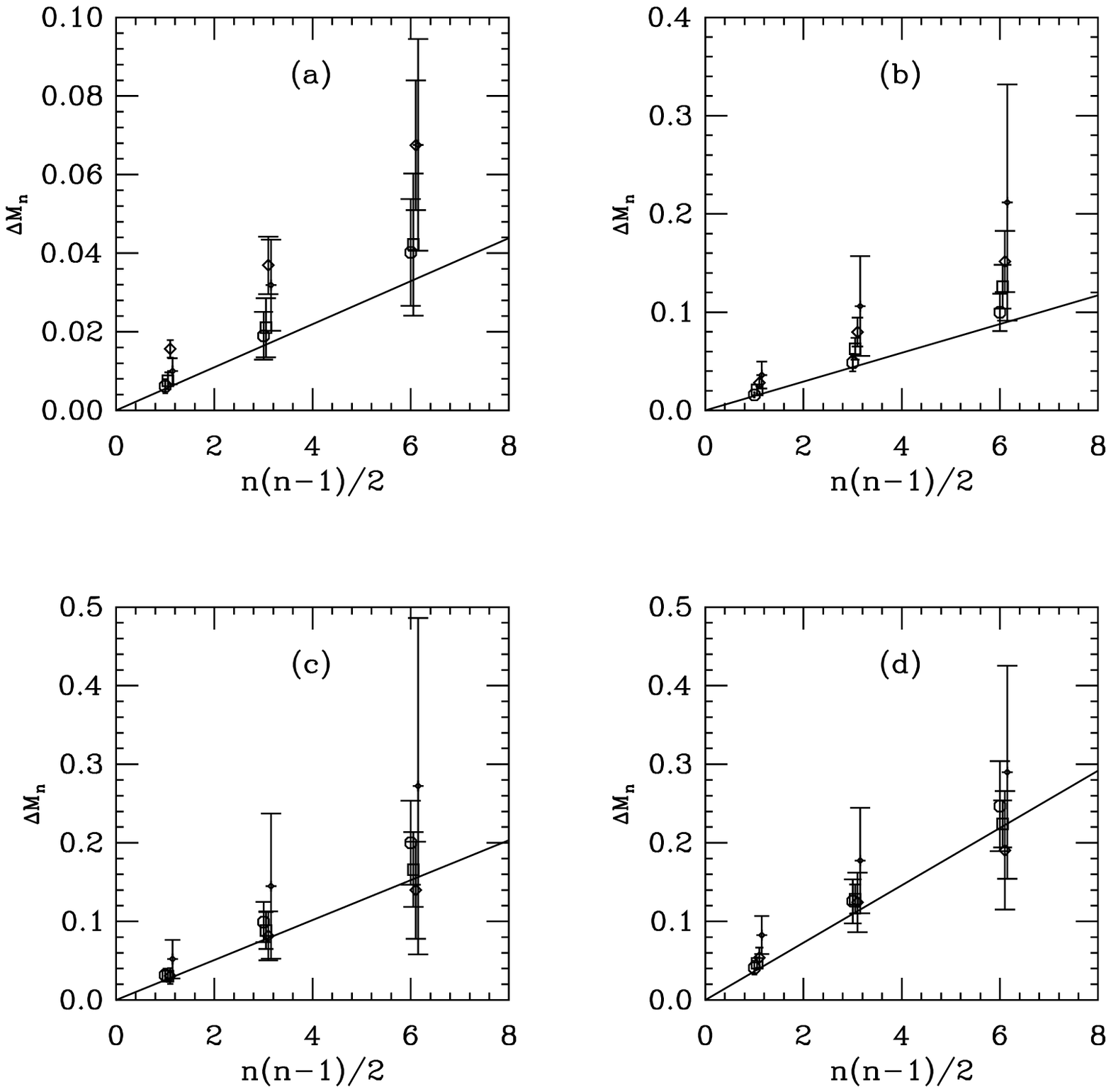}
\end{center}
\caption{Effective mass differences $\Delta M_n = nM_1 - M_n$ hadronic correlators
 in quenched $SU(3)$, $\beta=6.0175$, $\kappa=0.125$.
Symbols are octagons for $t=4$, squares for $t=6$, diamonds for $t=8$ and fancy diamonds for $t=10$.
The line is a fit to the slope of $\kappa_2$ for  $3\le t \le 8$.
(a) pseudoscalar;
(b) vector meson;
(c) proton;
(d) delta.
\label{fig:dmsu3}}
\end{figure}

Let's next test Eq.~\ref{eq:twobody}. I just take effective masses and, under a jackknife, compute
$\Delta M = nM - M_n$. This should be linear in $n(n+1)$, and the slope
 should be given by the part of $\kappa_2$
for $\log C(t)$ which is linear in $t$. Fig.~\ref{fig:dmsu3} shows this behavior quite nicely
for hadron correlators in $SU(3)$.
 The line is a fit to $S$ (see Eq.~\ref{eq:twobody}) over the range $3\le t \le 8$.

\begin{figure}
\begin{center}
\includegraphics[width=\textwidth,clip]{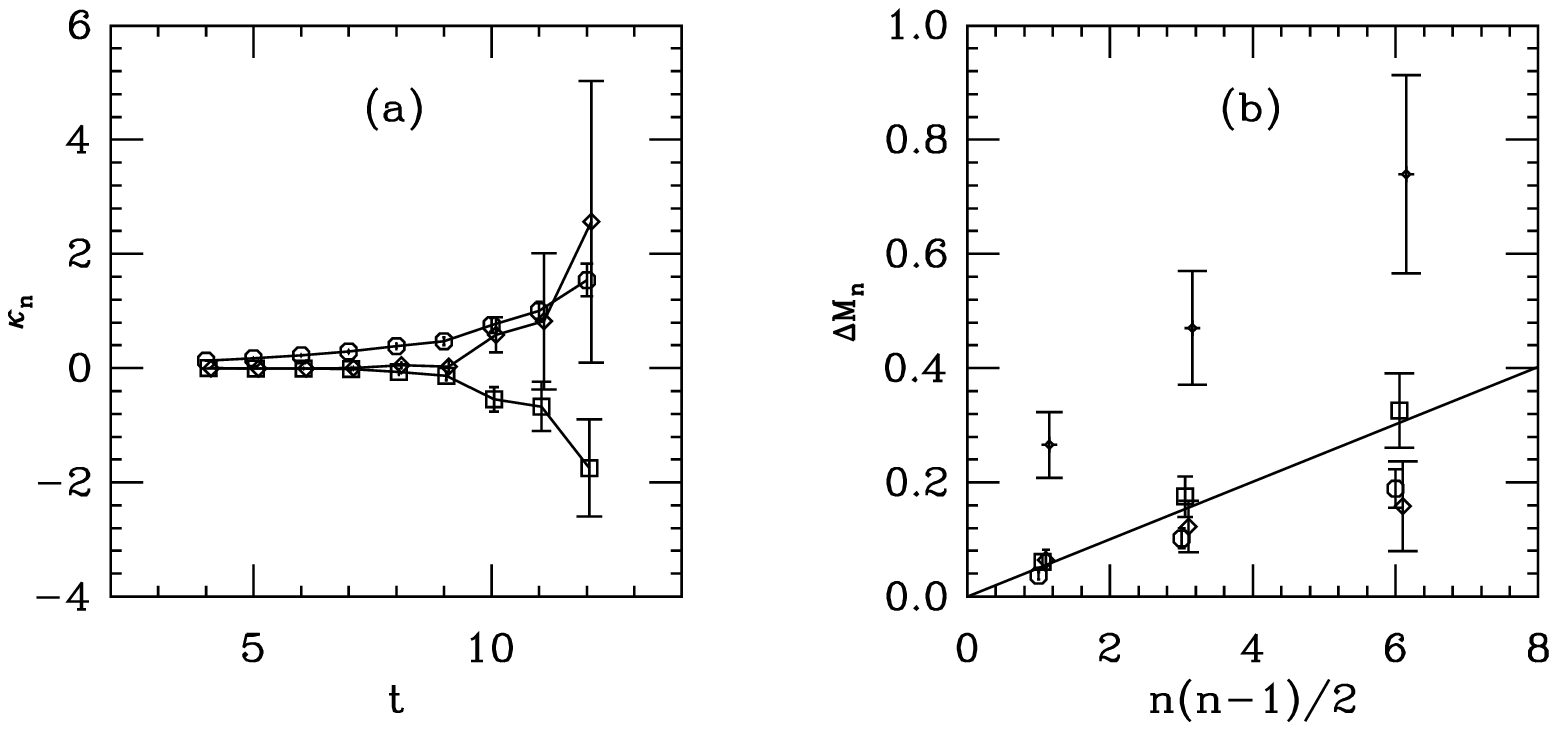}
\end{center}
\caption{(a) Cumulants and (b) mass differences for the $SU(5)$ $J=5/2$ baryon.
In panel (a), $\kappa_2$, $\kappa_3$ and $\kappa_4$ are shown by octagons, squares, and diamonds.
In panel (b), mass differences at $t=4$, 6, 8 , and 10 are shown as
octagons, squares, diamonds and fancy diamonds. The line is a fit to $S$.
\label{fig:diffdelt}}
\end{figure}

Recall panel (b) of Fig.~\ref{fig:moment}, showing the evolution of mass parameters at large $t$.
 Fig.~\ref{fig:diffdelt} shows  cumulants and the mass splitting for our $SU(5)$
$J=5/2$ state. Log-normal behavior works well at shorter $t$ and fails at the largest $t$.

\begin{figure}
\begin{center}
\includegraphics[width=0.9\textwidth,clip]{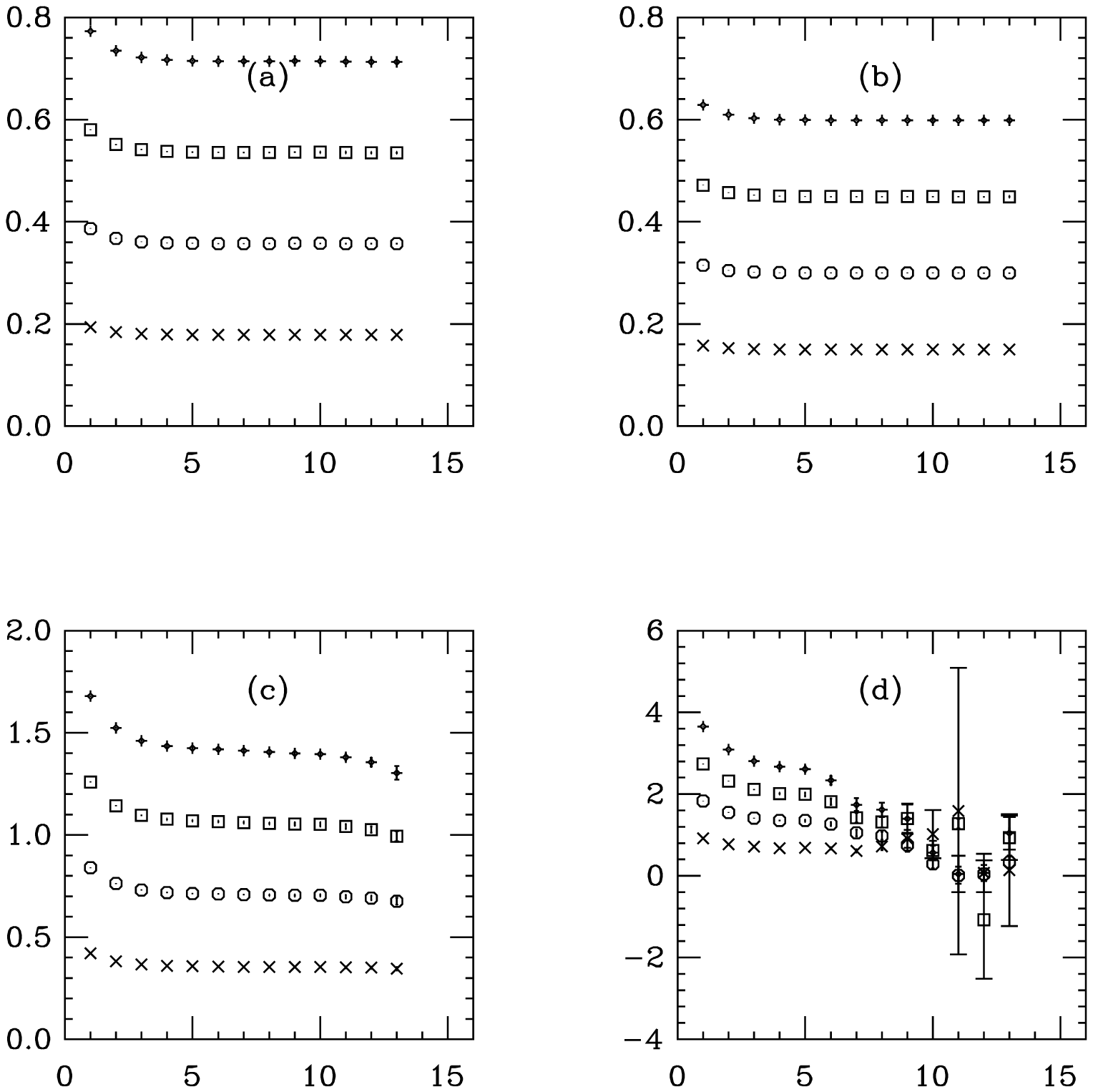}
\end{center}
\caption{Effective mass for moments of potentials in quenched $SU(3)$, $\beta=6.0175$.
Higher moments lie higher.
(a) $r=2$ planar loop;
(b) $\vec r= (1,1,1)$ loop;
(c) $r=3\sqrt{2}$ loop;
(d) $r=6\sqrt{2}$ loop.
\label{fig:mvr}}
\end{figure}

\begin{figure}
\begin{center}
\includegraphics[width=0.9\textwidth,clip]{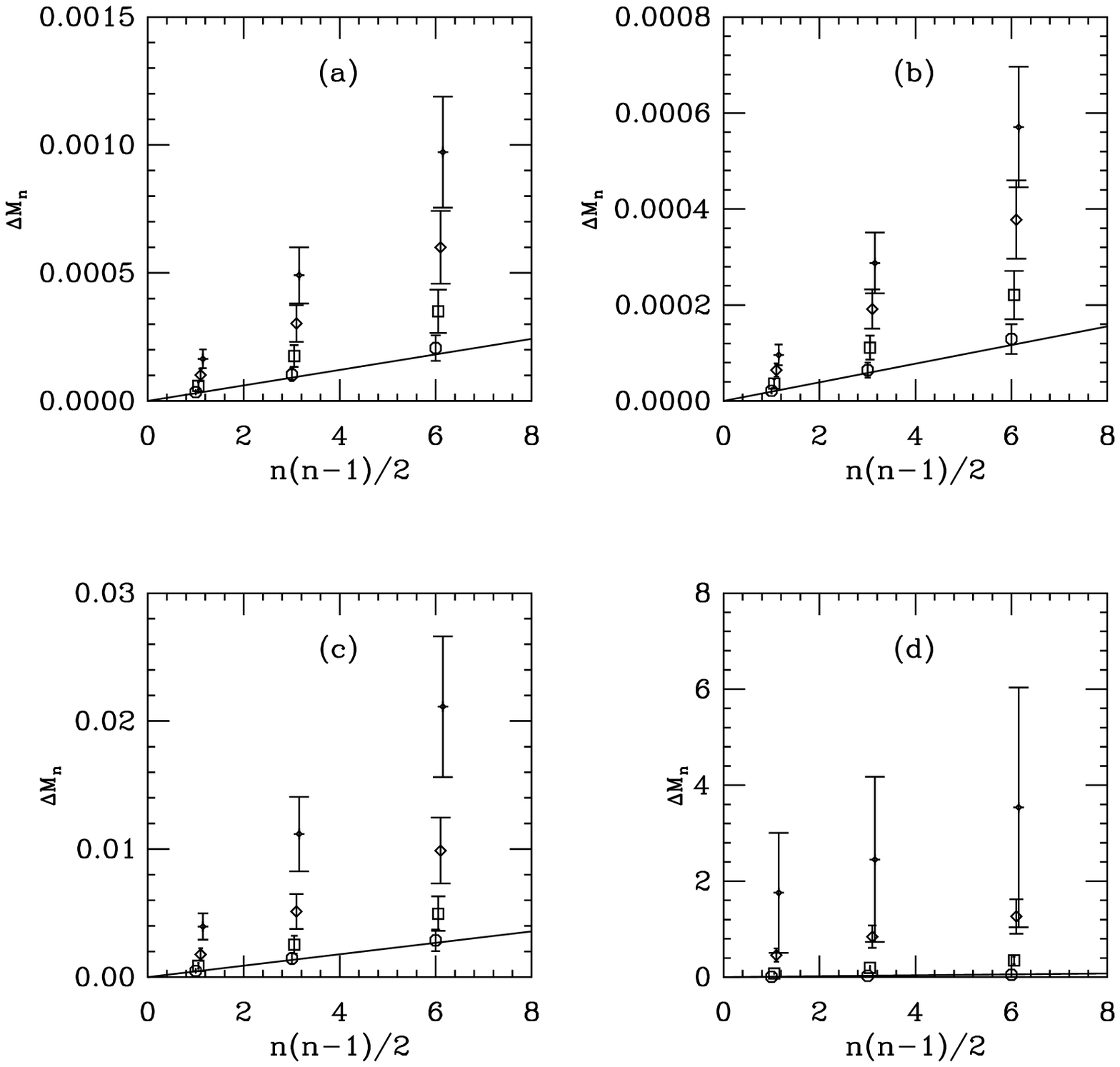}
\end{center}
\caption{Effective mass differences $\Delta M_n = nM_1 - M_n$ from Wilson loops in quenched $SU(3)$, $\beta=6.0175$.
Symbols are octagons for $t=4$, squares for $t=6$, diamonds for $t=8$ and fancy diamonds for $t=10$.
(a) $r=2$ planar loop;
(b) $\vec r= (1,1,1)$ loop;
(c) $r=3\sqrt{2}$ loop;
(d) $r=6\sqrt{2}$ loop.
\label{fig:dmvr}}
\end{figure}

Finally, we return to potentials. Figs.~\ref{fig:pot6.0175}, \ref{fig:mvr} and
\ref{fig:dmvr} show the consistency of log-normal behavior
(dominant $\kappa_2$, effective masses scaling as in Eq.~\ref{eq:twobody}) at short distances,
 and when the effective mass for
the moments falls, dominance of $\kappa_2$ goes away.

So to summarize: At small and intermediate $t$, hadronic correlators show log-normal behavior. This
is the $t$ range where the Lepage formula does not describe the effective mass of the moments of $C(t)$.
At large $t$, the Lepage formula does appear to describe the effective mass of the moments and correlators
cease to be log-normal.

\begin{figure}
\begin{center}
\includegraphics[width=0.9\textwidth,clip]{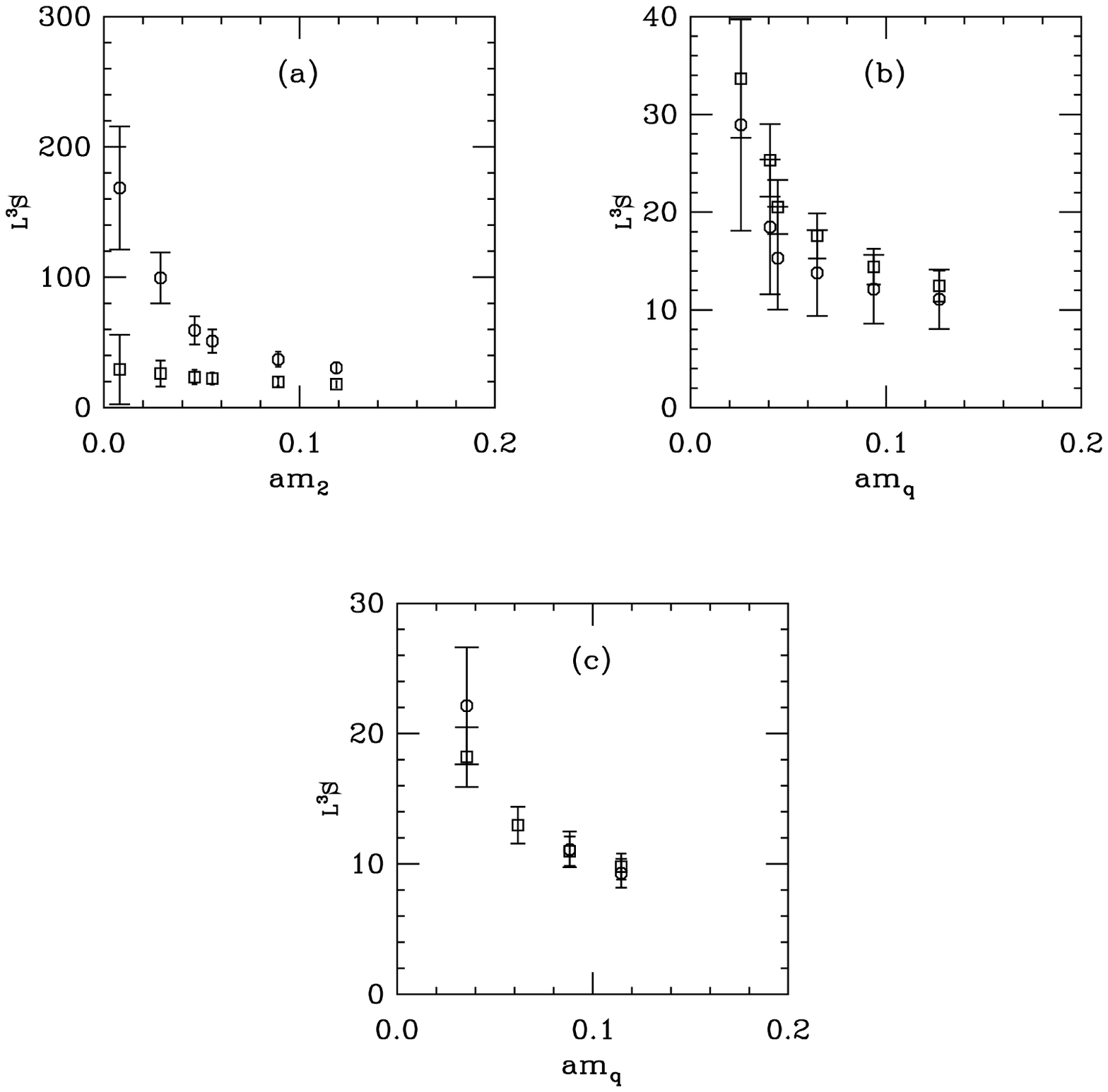}
\end{center}
\caption{Parameter $S$ from the slope of $\kappa_2$ with $t$, from Eq.~\protect{\ref{eq:S}}
for pseudoscalar correlators from quenched simulations. Squares $16^3$ volume; octagons, $12^3$ volume.
(a) $SU(3)$
(b) $SU(5)$
(c) $SU(7)$.
The $x$ axis is the AWI quark mass and all three data sets are matched in lattice spacing.
\label{fig:spseudo}}
\end{figure}

\begin{figure}
\begin{center}
\includegraphics[width=0.9\textwidth,clip]{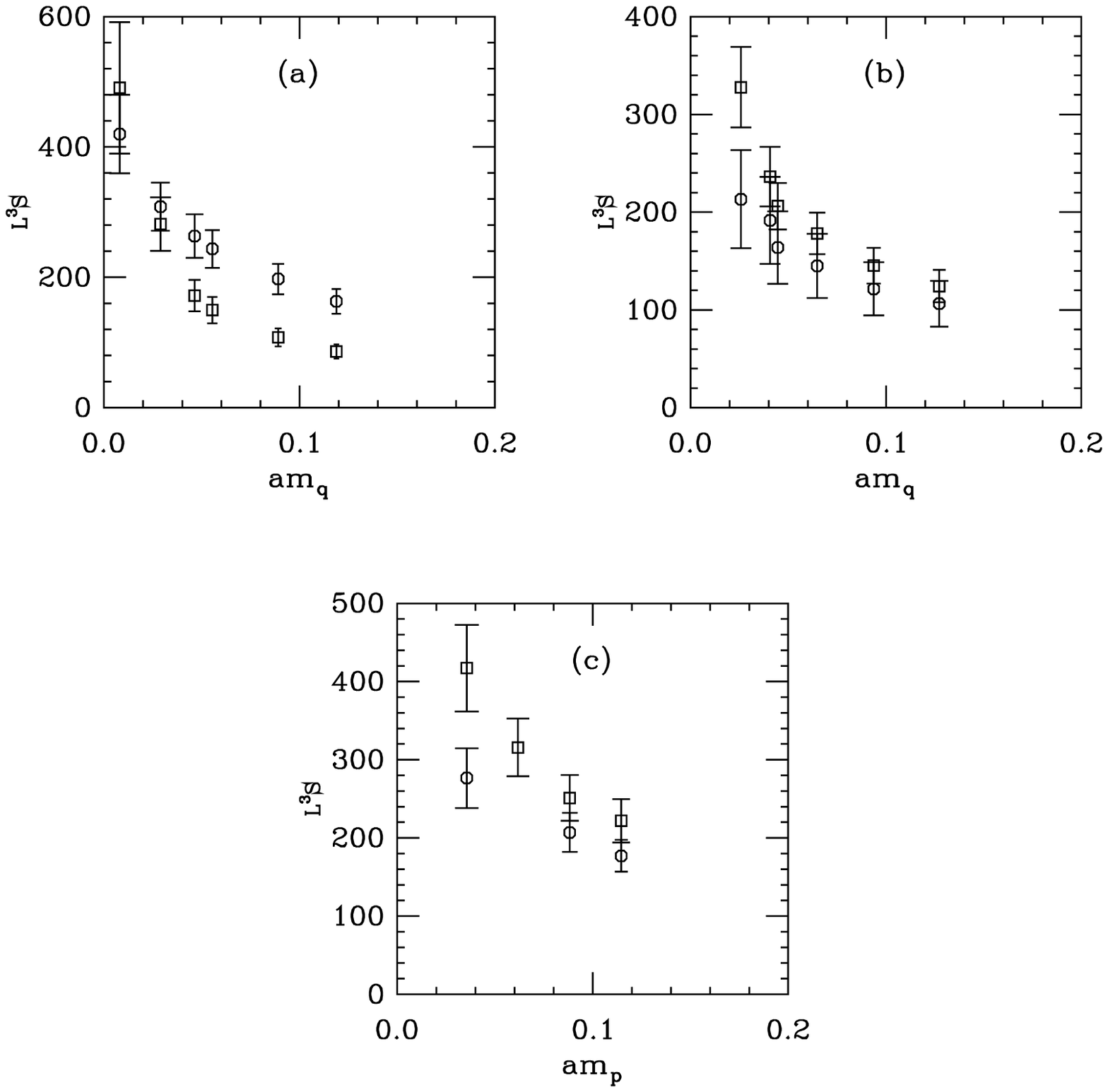}
\end{center}
\caption{Parameter $S$ from the slope of $\kappa_2$ with $t$, from Eq.~\protect{\ref{eq:S}}, scaled by the spatial simulation
volume,
for the highest-spin baryon from quenched simulations. Squares $16^3$ volume; octagons, $12^3$ volume.
(a) $SU(3)$
(b) $SU(5)$
(c) $SU(7)$.
The $x$ axis is the AWI quark mass and all three data sets are matched in lattice spacing.
\label{fig:sdelta2}}
\end{figure}

As a last observation, we can ask about volume dependence. I have two volumes for some of my quenched data sets.
 Figs.~\ref{fig:spseudo}-\ref{fig:sdelta2} show that the $S$ parameter, the slope of
 $\kappa_2$ with $t$, often scales inversely with the
simulation volume. I do not have enough other data sets to say more about this.

Are there any consequences of this observation? I can think of two.

First, the authors of Refs.
\cite{Endres:2011mm,Lee:2011sm,Endres:2011er,Endres:2011jm,Endres:2010sq,Nicholson:2010ms,Lee:2010qp}
have shown that, for their data sets, noisy signals can be tamed by replacing the average correlator by a
truncated cumulant sum,
\bee
\log \svev{C(t)} = \sum_{n=1}^N \frac{\kappa_n}{n!}.
\label{eq:replace}
\ee
Truncation of the sum at some finite $N$ introduces a systematic error on the mass, but it might
be less than the statistical error associated with averaging the original data set.
This might give a prediction for $M$ with a small statistical error. Varying $N$ and refitting would allow
an estimation of the systematic error.
Because the cumulants involve all the data, it would be necessary to fold this
procedure into a jackknife or bootstrap, and take the uncertainty in the fit parameters from 
the jackknife or bootstrap average.

An immediate problem doing this is that correlators at different time steps are themselves
strongly correlated. Usual fits take this correlation into account in the construction of the
correlation matrix for the chi-squared function. 
Information about time autocorrelations is lost when the cumulant sum is
performed time step by time step. If one is doing an effective mass fit with Eq.~\ref{eq:effmass},
 these correlations do not 
affect the mean value of  $M$ because the
fit
has no degrees of freedom. They do affect the uncertainties on the mass and intercept,
 but presumably a jackknife can handle that.
However, serious fits to lattice data are typically ``range fits''
over many values of $t$. Then correlations are important. The author has seen many fits which
miss the central values of the individual $C(t)$'s in an asymmetric way, due to the off-diagonal
correlations in the data.

\begin{figure}
\begin{center}
\includegraphics[width=0.8\textwidth,clip]{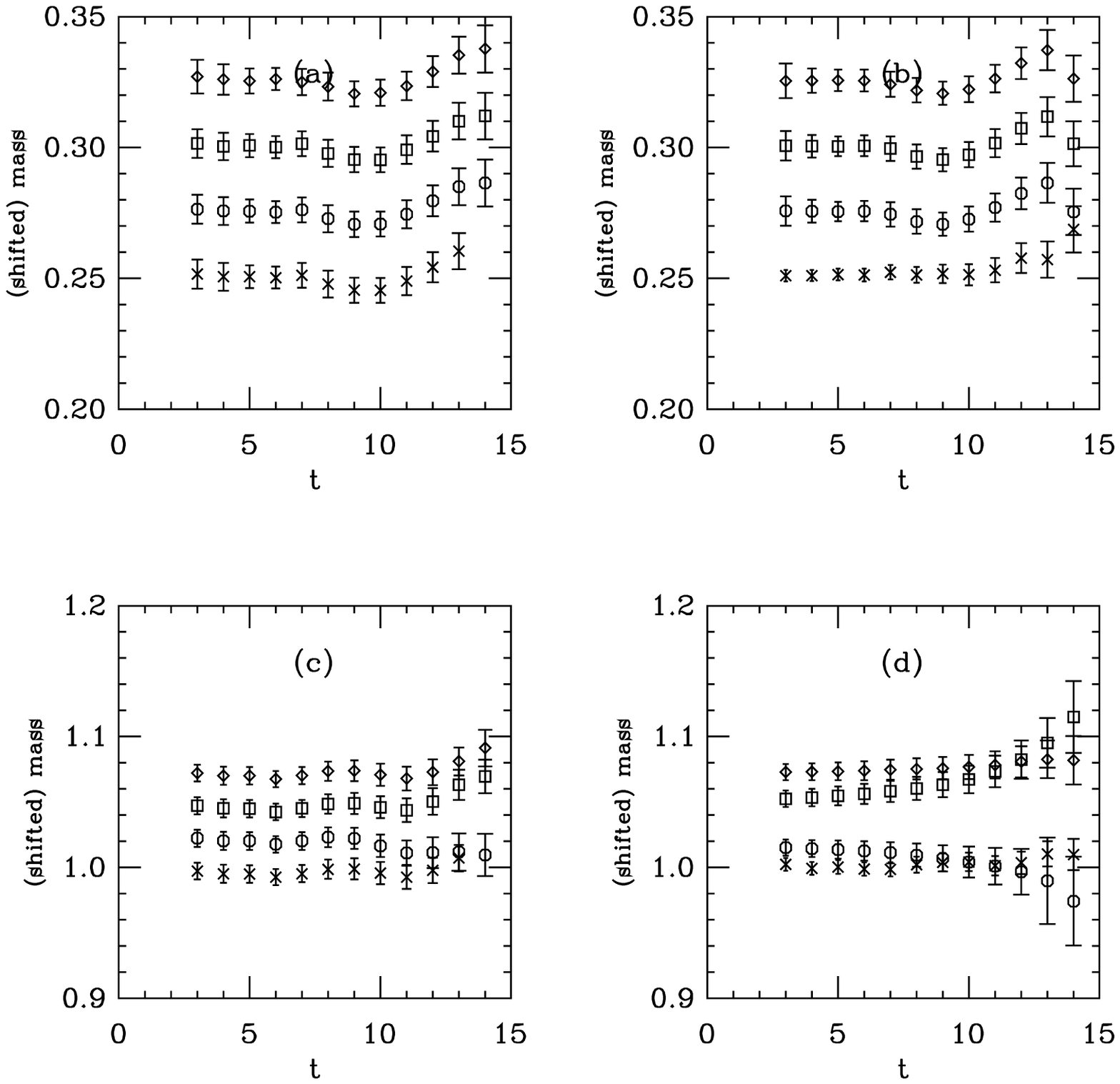}
\end{center}
\caption{Conventional and truncated cumulant fits to data, quenched $SU(3)$, $\beta=6.0175$.
Crosses show the conventional fit result. Octagons, squares and diamonds truncate the sum at $n=2$, 3, 4,
and the fit values are shifted vertically by multiples of 0.025.
Panels (a) and (b) show effective mass and range fits for the pseudoscalar, at   $\kappa=0.126$,
and panels (c) and (d), for the delta, at  $\kappa=0.123$.
\label{fig:tca}}
\end{figure}

For my quenched and dynamical data sets of meson and baryon correlators.
I compared conventional effective mass and range fits to fits where $C(t)$ was replaced by
a truncated cumulant sum. Even a truncation ending with $\kappa_2$ produced fit masses consistent
with the usual fits. However, unlike what
Refs.~\cite{Endres:2011mm,Lee:2011sm,Endres:2011er,Endres:2011jm,Endres:2010sq,Nicholson:2010ms,Lee:2010qp}
found, my uncertainties are not improved using the truncated cumulant sum.
I show results from a quenched $SU(3)$ example in Fig.~\ref{fig:tca}. Since the uncertainties are what
 I want to show, I offset the various orders of the truncated sum by constants. My observations are of 
course not a blanket statement that the truncated cumulant cannot be used to improve fits, only that I could not do it.

Second, one could take Eq.~\ref{eq:twobody}
seriously: the change in the width of the second cumulant measures a mass difference between
an $n$ hadron state in finite volume from $n$ times the single-hadron state.
(This connection was first made in the context of unitary Fermi gases by Nicholson \cite{Nicholson:2012zp}.)
In QCD, the mass difference gives information on a scattering length $a$:
\bee
\Delta M_n = \frac{4\pi a}{ML^3}
\label{eq:scattlength}
\ee
for particles in a box of volume $V$. Does the data support this?
We can test this hypotheses by taking the values of $S$ from
different volumes and just overlaying $L^3 S$. Examples were already shown, in
 Figs.~\ref{fig:spseudo} and \ref{fig:sdelta2}.
Sometimes the volume dependence is there.

Curious as it is, this connection cannot be made more precise in QCD. The $n$-hadron correlation functions
from which my masses are extracted (the moments) are composed of $n$ distinct color traces. These correlations
functions do not project onto a unique isospin, and so the right-hand side of Eq.~\ref{eq:scattlength}
 is -- at best --
a weighted sum of scattering lengths in different isospin channels.
And in QCD, three-body interactions have been measured by at least one group \cite{Detmold:2008fn}.
 They are not zero.

Since I do not have any crisp conclusions, I will finish with some questions.
First, is this behavior really ubiquitous?
It would be very interesting to look at distributions of correlation functions for 
two cases for which I don't have data:
One is simulation data from large lattices and small quark masses, where the lightest states in a generic channel
are not single-particle states, but multi-body ones. The rho channel when $m_\rho \gg 2m_\pi$ is an example.
Another example would be correlation functions of operators which are highly tuned (say 
from a variational calculation)
to project on a single state. I am also not satisfied with my comparisons of different volumes.

Second, if log-normality is there, why is it there? Since I see it in so many channels, it can't be a consequence
of the kind of correlator (baryon versus meson) or system (confining versus conformal).
Log-normal distributions commonly arise when an observable is a product of a set of
a set of independent positive random numbers.
 The variables in a lattice simulation of QCD
 are matrices, not numbers, and they are not completely random either -- the action weights the likelihood of
a configuration.
One thing that all the correlators I have examined have in common is that they involve
products of link matrices, and the number of link matrices involved in a correlator increases with
its $t$ value. This is certainly the case for Wilson loops. Hadronic correlators are built of
quark propagators. Quark propagators are themselves sums of products of link variables
connecting the source and the sink points of the propagator.
Because of the additive property
of cumulants, the cumulants of the log of the product of $\tau$ random variables are equal
to $\tau$ times the  cumulant of the individual distributions.  This certainly has the flavor of
the linear increase in $\kappa_2(t)$ observed in the data.

And finally, can log-normality be used to do anything quantitative?
So far, I have not been able to use it  to improve mass determinations along the lines of
\cite{Endres:2011mm,Lee:2011sm,Endres:2011er,Endres:2011jm,Endres:2010sq,Nicholson:2010ms,Lee:2010qp}.

As I said at the start, I am not sure whether approximate log-normality in lattice
correlator data is useful for anything. However,
I have to say: I have been looking at lattice data for a long time,
and it is quite curious to observe something new and (apparently) generic in it.

\begin{acknowledgments}
I would like to thank  M.~G.~Endres, D.~B.~Kaplan, J.~-W.~Lee and A.~N.~Nicholson
for much correspondence and conversation. Their Ref.~\cite{Endres:2011mm} directly inspired this investigation.
This work was supported  by the U.~S. Department of Energy.
\end{acknowledgments}



\begin{thebibliography}{99}
\bibitem{Detmold} The data was provided by W.~Detmold using results from the NPLQCD collaboration.
\end{thebibliography}
\end{document}